\documentstyle[12pt]{article}

\newcommand{\rf}[1]{(\ref{#1})}

\renewcommand{\thefootnote}{\fnsymbol{footnote}}

\newcommand{\newsection}{    
\setcounter{equation}{0}
\section}
\def\appendix#1{
  \addtocounter{section}{1}
  \setcounter{equation}{0}
  \renewcommand{\thesection}{\Alph{section}}
  \section*{Appendix \thesection\protect\indent \parbox[t]{11.715cm} {#1} }
  \addcontentsline{toc}{section}{Appendix \thesection\ \ \ #1}
  }

\def\nline{\,\nabla\kern -0.7em\raise0.2ex\hbox{/}\,\,}
\def\yline{\,y\kern -0.47em /}
\def\aline{\,a\kern -0.49em /}
\def\parline{\,\partial\kern -0.55em /\,\,}

\def\det{\hbox{det}}
\def\be{\begin{equation}}
\def\ee{\end{equation}}

\begin{document}

\begin{titlepage}

\vspace{.5cm}

\begin{center}
{\Large  \bf Super Poisson-Lie symmetry of the $GL(1|1)$ WZNW

\vspace{2mm}
model and worldsheet boundary conditions }\\[.2cm]
\vspace{1.1cm} {\large {A. Eghbali}{$^a$}${}^{}$\footnote{\
E-mail: a.eghbali@azaruniv.edu}
and A. Rezaei-Aghdam{$^b$}${}^{}$\footnote{\ Corresponding Author E-mail: rezaei-a@azaruniv.edu} }\\
\vspace{15pt}

{ {\footnotesize \it $^{a,b}$Department of Physics, Faculty of science, Azarbaijan Shahid Madani University,}
\\{\footnotesize \it 53714-161, Tabriz, Iran}
}\\

\end{center}

\vskip 0.6 cm

\begin{abstract}
We show that the WZNW model on the Lie supergroup $GL(1|1)$ has
super Poisson-Lie symmetry with the dual Lie supergroup ${B \oplus
A \oplus A_{1,1}}_|.i$. Then, we discuss about $D$-branes and
worldsheet boundary conditions on supermanifolds, in general, and
obtain the algebraic relations on the gluing supermatrix for the
Lie supergroup case. Finally, using the supercanonical
transformation description  of the super Poisson-Lie T-duality
transformation, we obtain formulae for the description of the dual
gluing supermatrix, then, we find the gluing supermatrix for the
WZNW model on
 $GL(1|1)$ and its dual model. We also discuss about different
 boundary conditions.
\end{abstract}

\end{titlepage}
\setcounter{page}{1}
\renewcommand{\thefootnote}{\arabic{footnote}}
\setcounter{footnote}{0}

\newsection{ \large {\bf Introduction}}
Field theory and two-dimensional sigma models with supermanifolds
as target space have recently received considerable attention,
because of their relations to condensed matter physics and
superstring models, respectively. As an example, the WZNW models
on supergroups are related to local logarithmic conformal field
theories  \cite{{schom},{Gotz},{Saleur}}. The first attempt in this
direction dates back to about 3 decades ago \cite{Henneaun}, where
the flat space GS superstring action was reproduced as a WZNW type
sigma model on the coset superspace ($\frac{D=10~ Poincare~
supergroup}{{SO(9,1)}}$). Then, this work is extended to the
curved background \cite{Tseytlin} and shown that type IIB
superstring on $AdS_5 \times S^5$ can be constructed from sigma
model on the coset superspace $\frac{SU(2,2|4)}{SO(4,1)\times
SO(5)}$. After then, superstring theory on $AdS_3 \times S^3$ is
related to WZNW model on $PSU(1,1|2)$ \cite{Berkovits} and also
superstring theory on $AdS_2 \times S^2$ is related to sigma
model on supercoset $\frac{PSU(1,1|2)}{U(1)\times U(1)}$
\cite{Bershadsky}. There are also other works in this direction,
see for instance \cite{Sorokin}.

\smallskip

On the other hand, T-duality is the most important symmetries of
string theory \cite{Busher}. Furthermore, Poisson-Lie T-duality, a
generalization of T-duality, does not require existence of
isometry in the original target manifold (as in usual T-duality)
\cite{{K.S1}, {K.S2}}. So, the studies of Poisson-Lie T-duality in
sigma models on supermanifolds and  duality in superstring
theories on AdS backgrounds are interesting problems. In the
previous works \cite{cosmology} we extended Poisson-Lie symmetry
to sigma models on supermanifolds and also constructed Poisson-Lie
T-dual sigma models on Lie supergroups \cite{ER}. In this paper we
show that the WZNW model on the Lie supergroup $GL(1|1)$ has super
Poisson-Lie symmetry with the dual Lie supergroup ${B \oplus A
\oplus A_{1,1}}_|.i$. In \cite{K.S22}, it was shown that the
mutually  $T$-dual sigma models on the cosets $D/G$ and $D/{\tilde
G}$ are the same, being equal to the WZNW model on ${ R}$; such
that ${\cal R}$ is the Lie subalgebra or $R$ directly is
identified with $D/G$ and $D/{\tilde G}$.  We note that until now
there is only one example \cite{K.S3} for mutually $T$-dual sigma
model on $G$  or ${\tilde G}$ such that one of the models is a
WZNW model. In that example, the WZNW model is a constraint model
because of singularity of the constant background matrix $E(e)$.
Here, we first show that the WZNW model  on the Lie supergroup
$GL(1|1)$ has super Poisson-Lie symmetry, then we obtain the
mutually $T$-dual sigma models on $GL(1|1)$ and its dual Lie
supergroup ${B \oplus A \oplus A_{1,1}}_|.i$. Note that the model
on Lie supergroup ${B \oplus A \oplus A_{1,1}}_|.i$ is not a WZNW
model, so it can not be conformal invariant. Furthermore, we
discuss about $D$-branes and worldsheet boundary conditions on
supermanifolds, in general, and specially  on Lie supergroups. For the
$GL(1|1)$ Lie supergroup, the $D$-brane and worldsheet boundary conditions previously have been studied
in \cite{creutzig1} (see, also \cite{creutzig2}). Here, we will present
new boundary conditions for this Lie supergroup and will study the effect of super Poisson-Lie T-duality
on these conditions.

The structure of this note is as follows. In Section 2 for self
containing of the paper and introducing the notations, we review
some aspects of the super Poisson-Lie T-dual sigma model on Lie
supergroup \cite{ER}. In Section 3, by using of direct calculation
of the super Poisson-Lie symmetry condition, we show that the WZNW
model on $GL(1|1)$ Lie supergroup has super Poisson-Lie symmetry
when the dual Lie supergroup is ${B \oplus A \oplus A_{1,1}}_|.i$.
Then, we obtain the mutually T-dual sigma model on the Drinfel'd
superdouble supergroup $(GL(1|1)\;,\;{ B \oplus A \oplus
A_{1,1}}_|.i)$ in Section 4; such that the original model is the
WZNW model on the Lie supergroup $GL(1|1)$. In Section 5, we first
discuss about $D$-branes and the
 worldsheet boundary conditions on supermanifolds and obtain general relation on the
gluing supermatrix which defines the relation between left- and right-movers on the worldsheet boundary; then we present algebraic form of these conditions when the target space is a Lie supergroup.
Then, using the supercanonical transformation description of the duality transformation, we derive a duality map for the gluing supermatrix which locally
defines the properties of the $D$-brane.
Finally, in the latter subsection of Section 5, we obtain the gluing supermatrices for the WZNW model on $GL(1|1)$ and its dual model as six cases.
Also, boundary conditions are discussed for the case 2.

\newsection{ \large{\bf Review of the super Poisson-Lie T-dual sigma models\\ ~~~on
supergroups}}
Let us start with a short review of the super Poisson-Lie symmetry
\cite{ER} on supermanifolds\footnote{Here we use the notation
presented by DeWitt's in \cite{D}.}. In what follows we shall
consider a nonlinear sigma model on a supermanifold $M$
as\footnote{Note that $G_{\Upsilon\Lambda}$ and
$B_{\Upsilon\Lambda}$ are supersymmetric metric and
antisupersymmetric tensor field, respectively, i.e.,
$$
G_{\Upsilon\Lambda}\;=\;(-1)^{|\Upsilon||\Lambda|}\;G_{\Lambda\Upsilon},~~~~~~~~~~~B_{\Upsilon\Lambda}\;=\;-(-1)^{|\Upsilon||\Lambda|}\;B_{\Lambda\Upsilon}.
$$
We will assume that the metric ${_{\Upsilon}G_{\Lambda}}$ is
superinvertible and its superinverse is denoted
$G^{\Upsilon\Lambda}$. In the above relations $|\Upsilon|$ denotes
the parity of $\Upsilon$, here and in the following we use the
notation \cite{D} $(-1)^{\Upsilon}:=(-1)^{|\Upsilon|}$.}
\begin{equation}\label{D1}
S\;=\;-\frac{T}{2}\int\!d\tau  d\sigma\;\left[ \sqrt{-h} h^{a b}
\partial_{a}\Phi ^{^{\Upsilon}}\; {\hspace{-0.5mm}_{_{\Upsilon}} {
G}}\hspace{-0.5mm}_{_{
\Lambda}}(\Phi)\;\partial_{b}\Phi^{^{\Lambda}}
+\epsilon^{ab}\partial_{a}\Phi{^{^\Upsilon}}\;
{\hspace{-0.5mm}_{_{\Upsilon}} { B}}\hspace{-0.5mm}_{_{
\Lambda}}(\Phi)\;\partial_{b}\Phi^{^{\Lambda}} \right],
\end{equation}
where $h_{ab}$  and  $\epsilon^{ab}$ are  the metric and
antisymmetric tensor on the worldsheet, respectively, such that
$h\equiv \det {h_{ab}}$ and the indices $a, b=\tau, \sigma$. The
coordinates $\Phi{^{^ \Upsilon}}$ include the bosonic coordinates
$X^{^{\mu}}$ and the fermionic ones $\Theta ^{^{\alpha}}$, and the
labels $\scriptsize{\Upsilon}$ and $\scriptsize {\Lambda}$ run
over $(\mu , \alpha)$. The labels $\mu$ and $\alpha$ run from $0$
to $d_B-1$ and from $1$ to $d_F$, respectively, such that $d_F$ is
an even number\footnote{For invertibility of the metric
$G_{\Upsilon\Lambda}$, $d_F$ must be even.}. We denote the
dimension of the bosonic directions by $d_B$ and the dimension of
the fermionic directions by $d_F$. Thus, the superdimension of the
supermanifold is written as $(d_B|d_F)$. One can write the action
\rf{D1} in lightcone coordinates and then obtain
\begin{equation}\label{sigma1}
S\;=\;\frac{1}{2}\int\!d\xi^{+}\wedge d\xi^{-}\;\partial_{+}\Phi
^{^{\Upsilon}}\; {\hspace{-0.5mm}_{_{\Upsilon}} {
{\cal{E}}}}\hspace{-0.5mm}_{_{
\Lambda}}(\Phi)\;\partial_{-}\Phi^{^{\Lambda}},
\end{equation}
where $\partial_{\pm}$ are the derivatives with respect to the
standard lightcone variables $\xi^{\pm} \;\equiv\;
\frac{1}{2}(\tau \pm \sigma)$ and ${{\cal {\cal
{E}}}}_{\Upsilon\Lambda}\;=\;G_{\Upsilon\Lambda}
+B_{\Upsilon\Lambda}$.

Now we  assume that the supergroup $G$ acts freely on $M$ from
right. The Hodge star of Noether's current one-forms corresponding
to the right action of the supergroup $G$ on the target $M$ of the
sigma model \rf{sigma1} has the following form
\begin{equation}\label{sigma3}
\star {_iJ}\;
=\;(-1)^{\Upsilon+\Lambda}\;{{_iV}^{(L,l)}}^{^{\Lambda}}
\;\partial_{+} \Phi ^{^{\Upsilon}}\;{\cal E}_{_{{\Upsilon\Lambda
}}}\;d\xi^{+} -(-1)^{\Lambda}\;{{_iV}^{(L,l)}}^{^{\Lambda}}\;{\cal
E}_{_{{\Lambda\Upsilon}}}\;\partial_{-} \Phi^{^{\Upsilon}} d\xi^{-},
\end{equation}
where ${{_iV}}^{^{\Upsilon}}$'s are the left invariant supervector
fields (defined with left derivative)\footnote{From now on we will
omit the superscripts $(L,l)$ on ${{_iV}^{(L,l)}}^{^\Upsilon}$. }. Now, we
demand that the  forms $\star {_iJ}$  on the extremal surfaces
$\Phi ^{^{\Upsilon}}(\xi^{+},\xi^{-})$ satisfy  the Maurer-cartan
equation \cite{D}
\begin{equation}\label{sigma4}
d \star{_iJ} \;= \;-(-1)^{jk}\;\frac{1}{2} {\tilde{f}^{jk}}_{\; \;
\; \; \; i} \star{_jJ} \wedge \star {_kJ},
\end{equation}
where ${\tilde{f}^{jk}}_{\; \; \; \; \; i}$ are are structure
constants of Lie superalgebra $\tilde {\bf \mathcal{G}}$ (the dual
Lie superalgebra to $\mathcal{G}$). Then, the condition of the
super Poisson-Lie symmetry for the sigma model \rf{sigma1} is
given by \cite{D}
\begin{equation}\label{sigma5}
{\cal L}_{V_i}({\cal E}_{_{{\Upsilon\Lambda }}})
=(-1)^{i(\Upsilon+k)}\;
 {\cal E}_{_{{\Upsilon\Xi
}}} {(V^{st})^{^{\Xi}}}_{k}\;(\tilde
{\cal{Y}}_i)^{kj}\;{{_jV}}^{^{\Omega}}\;{\hspace{-0.5mm}_{_{\Omega}}
{ {\cal{E}}}}\hspace{-0.5mm}_{_{ \Lambda}}
\end{equation}
where $(\tilde {\cal{Y}}_i)^{jk} = -{{\tilde f}_{\;\;\;\;i}}^{jk}$ are the adjoint
representations of Lie superalgebra $\mathcal{G}$ and ''$st$'' stands for the
{\it supertranspose} \cite{D}. As mentioned in \cite{ER}, the integrability condition for
Lie superderivative gives  compatibility between the structure constants of
 Lie superalgebras $ {\bf \mathcal{G}}$ and  $\tilde {\bf \mathcal{G}}$
which are well-known as the mixed super Jacobi identities of
 $({\bf \mathcal{G}} , \tilde {\bf \mathcal{G}})$ \cite{{ER1}}.
\begin{equation}\label{sigma6}
{f^k}_{ij}\;{\tilde{f}^{ml}}_{\;  \; \; \;
k}\;=\;(-1)^{il}\;{f^m}_{ik}\;{\tilde{f}^{kl}}_{\;  \; \; \; j}
+{f^l}_{ik}\;{\tilde{f}^{mk}}_{\;  \; \; \;
j}+{f^m}_{kj}\;{\tilde{f}^{kl}}_{\;  \; \; \; i}
+(-1)^{mj}\;{f^l}_{kj}\;{\tilde{f}^{mk}}_{\;  \; \; \; i}.
\end{equation}
In the following, we shall consider T-dual sigma model on
supergroup $G$. To this end, suppose $G$ acts transitively and
freely on $M$; then the target can be identified with the
supergroup $G$. Now we assume  $\varepsilon^{+}$ is an
$(d_B|d_F)$-dimensional  linear subsuperspace and
$\varepsilon^{-}$ is its orthogonal complement such that
$\varepsilon^{+}+\varepsilon^{-}$ span the Lie superalgebra
${\cal{D}}= ({\mathcal{G}} | \tilde {\bf \mathcal{G}})$, i.e.,
Drinfel'd superdouble\footnote {A Drinfel'd superdouble \cite{ER4}
is a Lie superalgebra ${\cal{D}}$ which decomposes into the direct
sum, as supervector spaces, of two maximally superisotropic Lie
subsuperalgebras ${\mathcal{G}}$ and $\tilde {\mathcal{G}}$, each
corresponding to a Poisson-Lie supergroup ($G$ and $\tilde G$),
such that the subsuperalgebras are duals of each other in the
usual sense, i.e., $\tilde {\mathcal{G}}={\mathcal{G}}^*$.}. To
determine a dual pair of the sigma models with the targets $G$ and
$\tilde G$, one can consider the following equation of motion for
the mapping $l(\xi^{+} , \xi^{-})$ from the worldsheet into the
Drinfel'd superdouble supergroup $D$ \cite{ER}
\begin{equation}\label{sigma7}
< \partial_{\pm}l l^{-1}\;,\;  \varepsilon^{\mp}  >\;=\;0,
\end{equation}
where $<.\;,\;.>$ means the invariant bilinear form on the
superdouble. Using the Eq. \rf{sigma7} and the decomposition of an
arbitrary element of $D$ in the vicinity of the unit element of
$D$ as
\begin{equation}\label{sigma8}
l(\xi^{+} , \xi^{-})\; =\; g(\xi^{+} , \xi^{-}) \tilde{h}(\xi^{+}
, \xi^{-}),\qquad  ~~~~ g\in G, \quad ~~ \tilde{h}\in \tilde{G},
\end{equation}
we obtain
\begin{equation}\label{sigma9}
<g^{-1} \partial_{\pm}g + \partial_{\pm}\tilde{h}\hspace{0.5mm}
\tilde{h}^{-1} ,g^{-1} \varepsilon^{\mp} g> \;=\; 0,
\end{equation}
for which
\begin{equation}\label{sigma10}
g^{-1} \varepsilon^{\pm} g\; = \;Span\{X_i \pm E^{\pm}_{ij}(g)
\tilde{X}^j\},
\end{equation}
such that $E^{-}_{ij}= (E^{+}_{ij})^{st} = (-1)^{ij}E^{+}_{ji}$;
$X_i$ and $\tilde X^i$ are basis of the respective Lie
superalgebras $ {\bf \mathcal{G}}$ and $\tilde {\bf \mathcal{G}}$.
It is crucial for super Poisson-Lie T-duality that the
superalgebras generated by $X_i$ and $\tilde X^i$ form a pair of
maximally superisotropic subsuperalgebras into the Drinfel'd
superdouble so that
$$
< X_i , X_j >\; =\; < {\tilde X}^i , {\tilde X}^j
>\;=\;0,~~~~~~~~~~
$$
\vspace{-4mm}
\begin{equation}\label{sigma11}
< X_i , {\tilde X}^j >\; =\;  (-1)^{ij} < {\tilde X}^j , X_i
>\;=\;(-1)^{i}{_i\delta} \hspace{1mm}^j.
\end{equation}
Now, one can write the action \rf{sigma1} in the following form
\begin{equation}\label{sigma12}
S\;=\;\frac{1}{2}\int\!d\xi^{+}\wedge
d\xi^{-}\;{R_{+}}{\hspace{-1mm}^{(l)^i}}\; {\hspace{-0.5mm}_i {
F^+}}\hspace{-2mm}_{
j}\;{^j{\hspace{-0.75mm}}R_{-}}^{{\hspace{-1mm}(l)}},
\end{equation}
where ${R_{\pm}}{\hspace{-1mm}^{(l)^i}}$'s are right invariant
one-forms with left derivative, i.e.,
\begin{equation}\label{sigma13}
{R_{+}}{\hspace{-1mm}^{(l)^i}}\;=\;\partial_{+} \Phi^{^{\Upsilon}}
\; {{_{_\Upsilon}} R}^{(l)^i}\;=\;(\partial_{+}g g^{-1})^i,
\end{equation}
\begin{equation}\label{sigma14}
{^j{\hspace{-1mm}}R_{-}}^{{\hspace{-1mm}(l)}}\; ={{\left(
{{R_-}^{\hspace{-1mm}(l)}}^{st}\right)}^j}{\;_{_{\Upsilon}}}\;
\partial_{-}
\Phi^{^{\Upsilon}} \;=\; (\partial_{-}g g^{-1})^j,
\end{equation}
and\footnote{Here, one must use of superdeterminant and
superinverse formulae \cite{D}.}
\begin{equation}\label{sigma15}
{ F^{+}(g)}\;=\;\Big(\Pi(g)+ ({E^{+}})^{-1}(e)\Big)^{-1},
\end{equation}
such that
\begin{equation}\label{sigma16}
\Pi^{ij}(g)\;=\;b^{ik}(g)\; {_k(a^{-1})}^j(g),
\end{equation}
where the matrices $a(g)$ and $b(g)$ are constructed using
\begin{equation}\label{sigma17}
g^{-1} X_i\; g\; =\;(-1)^j\;a_i^{\;\;j}(g)\; X_j,~~~~~~~~~
\end{equation}
\begin{equation}\label{sigma18}
~~~~~~g^{-1} \tilde{X}^i g\; =\; (-1)^j\;{b^{ij}(g)}\;{X_j} +
d^i_{\;\;j}(g)\; \tilde{X}^j,
\end{equation}
and consistency restricts them to obey
\begin{equation}\label{sigma19}
a(g^{-1}) \; =\; a^{-1}(g)\;=\;d^{st}(g),~~~~~~~~~~
\Pi(g)\;=\;-\Pi^{st}(g).
\end{equation}
We expect that there exists an equivalent T-dual sigma model in
which the roles of ${\bf \mathcal{G}}$ and $\tilde {\bf
\mathcal{G}}$ are exchanged. So, one can repeat all steps of the
previous construction to end up with the following T-dual sigma
model
\begin{equation}\label{sigma20}
\tilde S\;=\;\frac{1}{2}\int\ d\xi^{+}\wedge d\xi^{-}\;{\tilde {
R^{(l)}_{+}}\hspace{-0.5mm}_i}\; {{{{\tilde F}^{+ij}}}}
\;{_j{\hspace{-0.5mm}}{\tilde R}_{-}}^{{\hspace{-2mm}(l)}},
\end{equation}
where
\begin{equation}\label{sigma21}
\tilde { F^{+}}(\tilde g)\;=\;\Big(\tilde \Pi(\tilde g)+ {({{\tilde
E}^{+}})^{-1}(\tilde e) \Big)}^{-1}.
\end{equation}
Indeed, at the origin of the supergroup ($g=e$ and ${\tilde
g}=\tilde e$) the relation  between the matrices $E^{\pm}(e)$ and
${\tilde E}^{\pm}(\tilde e)$ are given by
\begin{equation}\label{sigma22}
E^{\pm}(e) {\tilde E}^{\pm}(\tilde e)\;=\;{\tilde E}^{\pm}(\tilde
e) E^{\pm}(e)\;=\;I.
\end{equation}

\newsection{\large {\bf Super Poisson-Lie symmetry of the $GL(1|1)$  WZNW model}}

The WZNW model based on supergroup $G$  takes the following
standard form
$$
S_{WZNW}(g) \; = \;  \frac{k}{4\pi} \int_{\Sigma} d\xi^{+} \wedge
d\xi^{-}\; < g^{-1} \partial_{+}g , \; g^{-1}
\partial_{-}g  >~~~~~~~~~~~~~
$$
\vspace{-5mm}
\begin{equation}\label{wz1}
\;\;\; +\; \frac{k}{24\pi} \int_{B} < g^{-1}dg\; \hat{,}
\;[g^{-1}dg \;\hat{,}\;g^{-1}dg]
 >,\
\end{equation}
where  the integrations are over worldsheet $\Sigma$ and a
3-dimensional manifold with boundary $\partial B = \Sigma$,
respectively, and $g^{-1} \partial_{\alpha}g $ are the left
invariant one-forms (with left derivative) on supergroup $G$ so
that they may be expressed as
\begin{equation}\label{wz3}
L^{(l)}_{\alpha}\;\equiv\;g^{-1}
\partial_{\alpha}g\;=\;(-1)^i (g^{-1} \partial_{\alpha}g)^i X_{i}.
\end{equation}
The WZNW action \rf{wz1} then can be rewritten in terms of the
$L^{\hspace{-0.5mm}(l)i}_{\alpha}$'s
\begin{equation}\label{wz4}
S_{WZNW}(g) =  \frac{k}{4\pi} \int_{\Sigma} d^2\xi\;
L^{\hspace{-0.5mm}(l)i}_{+}\;{_i\Omega}_j\;
L^{\hspace{-0.5mm}(l)j}_{-}-\frac{k}{24\pi} \int_{B} d^3 \xi
(-1)^{jk} \varepsilon^{ \gamma \alpha \beta}
L^{\hspace{-0.5mm}(l)i}_{\gamma}
\;{_i\Omega}_l\;L^{\hspace{-0.5mm}(l)j}_{\alpha}
({\cal{Y}}^l)_{jk}\; L^{\hspace{-0.5mm}(l)k}_{\beta},
\end{equation}
where $({\cal{Y}}^l)_{jk} = -{{f}^{l}}_{jk}$ are the adjoint
representations of Lie superalgebra $\mathcal{G}$ and
${\Omega}_{ij}\;=\;<X_i\;,\;X_j>=(-1)^{ij} {\Omega}_{ji}$ is
non-degenerate supersymmetric ad-invariant metric on
$\mathcal{G}$. Using the definition of metric ${\Omega}_{ij}$ and
ad-invariant  inner product on $\mathcal{G}$ as
\begin{equation}\label{wz5}
< X_i \; , \;[X_j , X_k] >\;=\;< [X_i , X_j] \; , \; X_k >,
\end{equation}
we find
\begin{equation}\label{wz6}
{\cal {X}}_i{\Omega}+({\cal {X}}_i{\Omega})^{st}\;=\;0,
\end{equation}
where $({{\cal{X}}_i)_{j}}^{\;k} = -{{f}_{ij}}^{\;k}$. Before
proceeding to write \rf{wz1} on the Lie supergroup $GL(1|1)$, let
us introduce the $gl(1|1)$ Lie superalgebra. The Lie superalgebra
$gl(1|1)$ has $(2|2)$-superdimension  with bosonic and fermionic
generators denoted by $H$, $Z$ \footnote{$Z$ is central generator,
i.e., it commutes with all other elements of $gl(1 |1)$.} and by
$Q_+$, $Q_-$, respectively. These four generators obey the
following set of non-trivial (anti)commutation relations \cite{B},
\cite{ER6},  \cite{schom}
\begin{equation}\label{wz7}
[H , Q_+]=Q_+,~~~~~~[H , Q_-]=-Q_-,~~~~~\{Q_+ , Q_-\}= Z.
\end{equation}
Here, we obtain a non-degenerate general solution to Eq. \rf{wz6}
as follows
\begin{equation}\label{wz7}
{\Omega}_{ij}\;=\; \pmatrix{b & a & 0 & 0 \cr a & 0 & 0 & 0 \cr 0
& 0 & 0 & a \cr 0 & 0 & -a & 0},\qquad   a\in \Re-\{0\},\;\;\; b\in \Re.
\end{equation}
In order to write \rf{wz1} explicitly, we need to find the
$L^{(l)}_{\alpha}$'s. To this purpose we use the following
parametrization of the Lie supergroup $GL(1|1)$ \cite{schom}:
\begin{equation}\label{wz8}
g\;=\; e^{\chi Q_-}  e^{y H+x Z}  e^{\psi Q_+}.
\end{equation}
The fields $x(\tau , \sigma)$ and $y(\tau , \sigma)$ are bosonic
while $\psi(\tau , \sigma)$ and $\chi(\tau , \sigma)$ are
fermionic. Inserting our specific choice of the parametrization
\rf{wz8}, the $L^{(l)}_{\alpha}$'s take the following form
\begin{equation}\label{wz9}
L^{(l)}_{\alpha}\;=\;\partial_{\alpha} y H +\partial_{\alpha} x
Z-\partial_{\alpha} \chi \;e^y \psi Z+ \partial_{\alpha} y \;\psi
Q_+ + \partial_{\alpha}\chi \;e^y Q_-,
\end{equation}
for which we can read off the $L^{\hspace{-0.5mm}(l)i}_{\alpha}$
and the terms that are being integrated over in \rf{wz4} are
calculated to be
\begin{equation}\label{wz10}
L^{\hspace{-0.5mm}(l)i}_{+}\;{_i\Omega}_j\;
L^{\hspace{-0.5mm}(l)j}_{-} \;=\;a[\partial_{+} y
\partial_{-} x + \partial_{+} x
\partial_{-} y -\partial_{+} \psi\; e^y
\partial_{-} \chi + \partial_{+} \chi\; e^y
\partial_{-} \psi],~~~~~~~~~
\end{equation}
\vspace{-4mm}
$$
(-1)^{jk}  L^{\hspace{-0.5mm}(l)i}_{\gamma}
\;{_i\Omega}_l\;L^{\hspace{-0.5mm}(l)j}_{\alpha}
({\cal{Y}}^l)_{jk}\; L^{\hspace{-0.5mm}(l)k}_{\beta}\;=\;-a
\partial_{\gamma} [-\partial_{\alpha} \psi e^y
\partial_{\beta} \chi -\partial_{\alpha} \chi e^y \partial_{\beta}
\psi~~~~~~~~~~~~~~~~~~~~~~~~~~~~
$$
\vspace{-4mm}
\begin{equation}\label{wz11}
~~~~~~~+\partial_{\alpha} e^y\;  \psi \partial_{\beta}
\chi-\partial_{\beta} e^y\;  \psi \partial_{\alpha}
\chi+\partial_{\alpha} e^y\;  \chi \partial_{\beta} \psi
-\partial_{\beta} e^y\;  \chi \partial_{\alpha} \psi].
\end{equation}
Finally, the $GL(1|1)$  WZNW action looks like
\begin{equation}\label{wz12}
S_{WZNW}(g) \; = \;  \frac{ak}{4\pi} \int_{\Sigma} d\xi^{+} \wedge
d\xi^{-}\;(\partial_{+} y
\partial_{-} x + \partial_{+} x
\partial_{-} y -2 \partial_{+} \psi\; e^y
\partial_{-} \chi).
\end{equation}
Here, we have assumed that $b=0$ in \rf{wz7}. One can derive the
$GL(1|1)$ WZNW action deduced in \cite{{schom}, {creutzig2}} by choosing $a=-1$.
On the other hand, by rescaling $a$ to $-\frac{2\pi}{k}$ and using
integrating by parts, the action \rf{wz12} is reduced to
$$
S_{WZNW}(g) \; = \;  \frac{1}{2} \int  d\xi^{+} \wedge
d\xi^{-}\;(-\partial_{+} y
\partial_{-} x -\partial_{+} x
\partial_{-} y + \partial_{+} \psi\; e^y
\partial_{-} \chi~
$$
\vspace{-3mm}
\begin{equation}\label{wz13}
~~~~~~~~~- \partial_{+} y\; e^y \psi
\partial_{-} \chi- \partial_{+} \chi\; e^y \psi
\partial_{-} y - \partial_{+} \chi\; e^y
\partial_{-} \psi).
\end{equation}
By regarding this action as a sigma model action of the form
\rf{sigma1}, we can read off the background matrix as follows:
\begin{equation}\label{wz14}
{\cal {E}}_{\Upsilon \Lambda}\;=\; \pmatrix{0 & -1 & 0 & -\psi e^y
\cr -1 & 0 & 0 & 0 \cr 0 & 0 & 0 & -e^y \cr \psi e^y & 0 & e^y &
0}.
\end{equation}
In the following, we shall investigate that the $GL(1|1)$ WZNW
model has super Poisson-Lie symmetry. To this end, we need the
left invariant supervector fields (with left derivative) on the
$GL(1|1)$. Substituting
$L^{(l)i}={\overrightarrow{d}}\Phi^{^{\Upsilon}}
{{_{_{\Upsilon}}}L}^{(l)^{i}}$ and ${_iV}= {_iV}{^{^{\Upsilon}}}
\frac{ \overrightarrow{\partial}}{\partial \Phi{^{^{\Upsilon}}}}$
into $<{_iV}\;,\; L^{(l)j}>\;=\;{_i{\delta}}^j$ we then obtain
\cite{ER}
\begin{equation}\label{wz15}
{_iV}^{^{\Upsilon}}\;=\;({_{\Upsilon}L}^{(l){i}})^{-1}.
\end{equation}
Thus, using the Eqs. \rf{wz9} and \rf{wz15}, ${_iV}$'s take the following form
$$
{_{_H}V}\;=\;\frac{ \overrightarrow{\partial}}{\partial  y}-\psi \frac{ \overrightarrow{\partial}}{\partial  \psi},
$$
\vspace{-3mm}
$$
{_{_Z}V}\;=\;\frac{ \overrightarrow{\partial}}{\partial  x},~~~~~~~~~~
$$
\vspace{-3mm}
$$
{_{_{Q_+}}V}\;=\;-\frac{ \overrightarrow{\partial}}{\partial \psi},~~~~~~~~
$$
\vspace{-3mm}
\begin{equation}\label{wz16}
~~~~~{_{_{Q_-}}V}\;=\;-\psi \frac{ \overrightarrow{\partial}}{\partial  x}-e^{-y}\frac{ \overrightarrow{\partial}}{\partial \chi}.
\end{equation}
Now, putting the relations \rf{wz14} and \rf{wz16} on the right
hand side of Eq. \rf{sigma5} and by direct calculation of the Lie
superderivatie corresponding to the ${_iV}$ of ${\cal
{E}}_{_{\Upsilon\Lambda}}$ \cite{ER}, then one can find the
non-zero structure constants of the dual pair to the $gl(1|1)$ Lie
superalgebra as
\begin{equation}\label{wz17}
{{\tilde f}_{\;\;\;\;3}}^{23}\;=\;-{{\tilde f}_{\;\;\;\;3}}^{32}\;=\;1.
\end{equation}
In our recent work \cite{ER6},  we classified all the dual Lie
superalgebras to the $gl(1|1)$. In \cite{ER6} the dual Lie
superalgebra \rf{wz17} has been labeled to the ${{\cal B} \oplus
{\cal A } \oplus {\cal A}_{1,1}}_|. i$, such that this  Lie
superalgebra is a decomposable Lie superalgebra of the type $(2 |
2)$ where is generated by the set of bosonic generators ${\tilde
X}^1=\tilde H, {\tilde X}^2=\tilde Z$ and fermionic ones ${\tilde
X}^3=\tilde Q_+, {\tilde X}^4=\tilde Q_-$ with the commutation
relations of \rf{wz17}. In the next section, we will show that the
original T-dual sigma model on the Drinfel'd superdouble
supergroup $(GL(1|1)\;,\;{ B \oplus A \oplus A_{1,1}}_|.i)$ is
equivalent to the $GL(1|1)$ WZNW model.

\vspace{4mm}

\newsection{\large {\bf Super Poisson-Lie dualizable sigma model on\\
the $(GL(1|1)\;,\;{  B \oplus A \oplus A_{1,1}}_|.i)$}}

In this section we will first introduce the Drinfel'd superdouble
generated by  the $gl(1|1)$ Lie superalgebra and it's dual $\tilde
{\mathcal{G}} = {{\cal B} \oplus {\cal A } \oplus {\cal
A}_{1,1}}_|. i$. We construct in particular, super Poisson-Lie
dualizable sigma models on the $GL(1|1)$ and ${  B \oplus A \oplus
A_{1,1}}_|.i$. The Manin supertriple $(gl(1|1)\;,\; {{\cal B}
\oplus {\cal A } \oplus {\cal A}_{1,1}}_|. i)$ possesses four
bosonic generators and four fermionic ones. We shall denote the
bosonic generators by $\{H, Z, \tilde H, \tilde Z\}$ and use
$\{Q_+, Q_-, \tilde Q_+, \tilde Q_-\}$ for fermionic generators.
The relations between these elements are given by \cite{ER6}
$$
[H , Q_+]=Q_+,~~~~~~~~[H , Q_-]=-Q_-,~~~~~~~\{Q_+ , Q_-\}=Z,
$$
\vspace{-4mm}
$$
[\tilde Z , \tilde Q_+]=\tilde Q_+,~~~~~~~~[H , \tilde
Q_+]=-\tilde Q_+,~~~~~~~[H , \tilde Q_-]=\tilde Q_-,
$$
\vspace{-4mm}
$$
~~~~~~[\tilde Z , Q_-]=\tilde Q_+,~~~~~~\{Q_- , \tilde
Q_-\}=\tilde H,~~~~~~~\{Q_+ , \tilde Q_+\}=Z-\tilde H,
$$
\vspace{-4mm}
\begin{equation}\label{Lie1}
[\tilde Z , Q_+]=-Q_+ +\tilde
Q_-.~~~~~~~~~~~~~~~~~~~~~~~~~~~~~~~~~~~~~~~~~~~~~\;
\end{equation}
In addition, the elements $Z$ and $\tilde H$ are central.

\vspace{1mm}

\subsection{ {\it The original model}}

There exist various choices that come with different
parametrizations of the Lie supergroup $GL(1|1)$. The convenient
parametrization for us is the same of \rf{wz8}. Using the
parametrization  \rf{wz8} and the relation \rf{sigma13} we have
explicitly
$$
{{R}^{(l)}_\pm}^H\;=\;{\partial_\pm}y,~~~~~~~~~~~~~~
$$
$$
{{R}^{(l)}_\pm}^Z\;=\;{\partial_\pm}x+{\partial_\pm}\psi e^y
\chi,~
$$
$$
{{R}^{(l)}_\pm}^{Q_+}\;=\;-e^y {\partial_\pm}\psi,~~~~~~~~~
$$
\begin{equation}\label{Lie2}
{{R}^{(l)}_\pm}^{Q_-}\;=\;-\chi {\partial_\pm}y-
{\partial_\pm}\chi.~~
\end{equation}
By a direct application of formulae \rf{sigma17} and \rf{sigma18}
the super Poisson structure is work out as follows:
\begin{equation}\label{Lie3}
\Pi^{ij}(g)\;=\;\pmatrix{0 & 0 & 0 & 0 \cr 0 & 0 & \psi e^y & 0
\cr 0 & -\psi e^y & 0 & 0 \cr  0 & 0 & 0 & 0 }.
\end{equation}
Then, choosing the inverse sigma model matrix $(E^+)^{-1}(e)$ at
the unit element of $GL(1|1)$ as
\begin{equation}\label{Lie4}
({E^+}^{-1})^{ij}(e)\;=\;\pmatrix{0 & -1 & 0 & 0 \cr -1 & 0 & 0 &
0 \cr 0 & 0& 0 & -1 \cr  0 & 0 & 1 & 0 },
\end{equation}
and finally, using the relations \rf {sigma15} and \rf{Lie2}, the
original model action \rf{sigma12} is obtained to be of the form
$$
S \; = \;  \frac{1}{2} \int  d\xi^{+} \wedge
d\xi^{-}\;(-\partial_{+} y
\partial_{-} x -\partial_{+} x
\partial_{-} y + \partial_{+} \psi\; e^y
\partial_{-} \chi~
$$
\vspace{-3mm}
\begin{equation}\label{Lie5}
~~~~~~~~~- \partial_{+} y\; e^y \psi
\partial_{-} \chi- \partial_{+} \chi\; e^y \psi
\partial_{-} y - \partial_{+} \chi\; e^y
\partial_{-} \psi).
\end{equation}
By identifying  the above action with the sigma model of the form
\rf{sigma1}, one can read off the background  supersymmetric
metric $G_{\Upsilon\Lambda}$ and antisupersymmetric tensor field
$B_{\Upsilon\Lambda}$ as follows:
\begin{equation}\label{Lie6}
G_{\Upsilon\Lambda}\;=\;\pmatrix{0 & -1 & 0 & 0 \cr -1 & 0 & 0 & 0
\cr 0 & 0& 0 & -e^y \cr  0 & 0 & e^y & 0
},~~~~~~~~~~~B_{\Upsilon\Lambda}\;=\;\pmatrix{0 & 0 & 0 & -\psi
e^y \cr 0 & 0 & 0 & 0 \cr 0 & 0& 0 & 0 \cr  \psi e^y & 0 & 0 & 0
}.
\end{equation}
This result is identical to the  conclusion of \rf{wz13}. Thus, we
showed that the original T-dual sigma model on the Drinfel'd
superdouble supergroup  $(GL(1|1)\;,\;{  B \oplus A \oplus
A_{1,1}}_|.i)$ is equivalent to the $GL(1|1)$ WZNW model.

\vspace{1mm}
\subsection{ {\it The dual model}}

In the same way to construct the dual model on the Lie supergroup
${ {B \oplus A \oplus A_{1,1}}_|}.i$, we use of the following
parametrization
\begin{equation}\label{Lie7}
\tilde g\;=\; e^{\tilde {\chi} {\tilde Q_-}}  e^{ {\tilde y}
{\tilde  H}+{\tilde x} {\tilde Z}} e^{{\tilde  \psi}{\tilde
Q_+}}.
\end{equation}
By using the \rf{Lie7} we find
\begin{equation}\label{Lie7-1}
\partial_+ \tilde g {\tilde g}^{-1}\;=\; {\partial_+}{\tilde y} \tilde H+ {\partial_+}{\tilde x} \tilde Z +
{\partial_+}{\tilde \psi} e^{\tilde x} {\tilde
Q}_++{\partial_+}{\tilde \chi}  {\tilde Q}_-,
\end{equation}
from which we can read off the ${{{\tilde R}_{+i}}^{(l)}}$'s  and
compute
\begin{equation}\label{Lie28}
{\tilde \Pi}_{ij}(\tilde g) \;=\;\pmatrix{0 & 0 & -e^{\tilde x}
{\tilde \psi} & {\tilde \chi} \cr 0 & 0 & 0 & 0 \cr e^{\tilde x}
{\tilde \psi} & 0 & 0 & 1-e^{\tilde x} \cr  -{\tilde \chi}  & 0 &
1-e^{\tilde x} & 0 }.
\end{equation}
Finally, using the Eq. \rf{sigma21} and the condition
\rf{sigma22}, the dual model action will have the following form
$$
\tilde S \; = \;  \frac{1}{2} \int  d\xi^{+} \wedge
d\xi^{-}\;\Big[-\partial_{+} {\tilde y}
\partial_{-} {\tilde x} -\partial_{+} {\tilde x}
\partial_{-} {\tilde y} +\frac{2}{e^{{\tilde x}}-2}( -\partial_{+} {\tilde x} \; \tilde \psi\; \tilde \chi\; \partial_{-} {\tilde x}
$$
\vspace{-3mm}
\begin{equation}\label{Lie5}
~~~~~~~~~+\partial_{+} {\tilde x} \; \tilde \psi \partial_{-}{
\tilde \chi}+\partial_{+}  {\tilde \psi} \;{\tilde \chi}
(e^{{\tilde x}}-1)\; \partial_{-} {\tilde x}-\partial_{+} {\tilde
\psi}
\partial_{-}{ \tilde \chi})-2 \partial_{+}{
\tilde \chi} \;{\tilde \psi} \;\partial_{-} {\tilde x}\Big].
\end{equation}

\newsection{\large {\bf $D$-branes  on supermanifolds and worldsheet boundary
conditions  }}
To study $D$-branes on supermanifolds, we impose the boundary
conditions for the bosonic and fermionic coordinates. Therefore,
$D$-branes on supermanifolds have more formations than those on
manifolds. For investigating  $D$-branes, we must study the
boundary conditions on the worldsheet. The open string may either
move about freely, in which case its ends obey Neumann boundary
conditions; or the ends of the string may be confined to a
subsuperspace, corresponding to Dirichlet conditions. One must
impose the Dirichlet or the Neumann conditions for each bosonic on
the boundary. Note that for supermanifolds, since the two fields
$\Theta^{\alpha}$ and $\Theta^{\alpha+1}$ become evident in pairs
in the action \cite{flat}, so one must more careful study the
fermionic parts $\Theta^{\alpha}$. Furthermore, since the
following conditions must be satisfied on the boundary
\begin{equation}\label{brane1}
\delta \Theta^{\alpha} \partial_{\sigma}
\Theta^{\alpha+1}=0,~~~~~~~~~~~\delta \Theta^{\alpha+1}
\partial_{\sigma} \Theta^{\alpha}=0,
\end{equation}
so, we need to impose the same boundary conditions for each pair
of fermionic directions. Thus,  the boundary conditions require
that $\delta \Theta^{\alpha}=\delta \Theta^{\alpha+1}=0$ or
$\partial_{\sigma} \Theta^{\alpha}=\partial_{\sigma}
\Theta^{\alpha+1}=0$. If the numbers of the directions with the
Neumann conditions be presented by $p+1$ for the bosonic
directions and $r$ for the fermionic directions, then, $r$ must be
an even number. $D$-branes with these configurations are called
$D{p|r}$-branes \cite{flat}. Consider a $(d_B|d_F)$-dimensional
target space with $D{p|r}$-branes, i.e., there are $d_B-(p +1)$
Dirichlet directions along which the field $X^{\mu}$ is frozen
($\partial_{0} X^{a} = 0, a = p + 1, ..., d_B-1$). At any given
point on a $Dp$-brane we can choose local coordinates such that
$X^a$ are the directions normal to the brane and $X^{m}$ ($m = 0,
..., p$) are coordinates on the brane. But, for the fermionic
part, we have $d_F-r$ Dirichlet directions, where $d_F-r$ is an
even number.


\subsection{ {\it Worldsheet boundary conditions}}

The worldsheet boundary is by definition confined to a $D$-brane.
Since the boundary relates left-moving fields $\partial_{+}
\Phi^{^{\Upsilon}}$ to the right-moving fields $\partial_{-}
\Phi^{^{\Upsilon}}$, we make a general ansatz for this relation.
The goal is then to find the restrictions on this ansatz arising
from varying the action \rf{D1}. The most general local boundary
condition may be expressed as\footnote{We note that the conditions
\rf{brane2}, \rf{brane9}, \rf{brane12} and \rf{brane13} are a
generalization of the bosonic conditions in \cite{CC}.}
\begin{equation}\label{brane2}
\partial_{-}\Phi^{^{\Upsilon}}\;=\;{{\cal
R}^{^{\Upsilon}}}_{{\;\Lambda}}(\Phi)\; \partial_{+}
\Phi^{^{\Lambda}},
\end{equation}
where $R^{^{\Upsilon}}_{{\;\;\Lambda}}$ is a locally defined
object which is called the {\it gluing supermatrix}. Now we assume
that $R^{^{\Upsilon}}_{{\;\;\Lambda}}$ is  in the form of a $2
\times 2$ block matrix as
\begin{equation}\label{brane3}
{{\cal R}^{^{\Upsilon}}}_{{\;\Lambda}}(\Phi)\;=\;\left(
\begin{tabular}{c|c}
                 ${{\cal
R}^{^{\mu}}}_{{\nu}}$ & ${{\cal R}^{^{\mu}}}_{{\beta}}$ \\
\hline
                 ${{\cal
R}^{^{\alpha}}}_{{\nu}}$ & ${{\cal
R}^{^{\alpha}}}_{{\beta}}$ \\
                 \end{tabular} \right),
\end{equation}
where the elements of the submatrices ${{\cal
R}^{^{\mu}}}_{{\nu}}$ and ${{\cal R}^{^{\alpha}}}_{{\beta}}$ are
$c$-numbers, while the elements of the submatrices ${{\cal
R}^{^{\mu}}}_{{\beta}}$ and ${{\cal R}^{^{\alpha}}}_{{\nu}}$ are
$a$-numbers.

These boundary conditions have to preserve conformal invariance at
the boundary. We know that each symmetry corresponds to a
conserved current, obtained by varying the action with respect to
the appropriate field. In the case of conformal invariance, the
corresponding current is the stress energy-momentum tensor and is
derived by varying the action \rf{D1} with respect to the metric
$h_{ab}$. Its components in lightcone coordinates are
\begin{equation}\label{brane6}
T_{\pm\pm}\;=\;(-1)^{\Upsilon}
\partial_{\pm} \Phi^{^{\Upsilon}}\;
G_{_{ \Upsilon \Lambda}}(\Phi)\;\partial_{\pm}\Phi^{^{\Lambda}}.
\end{equation}
The $T_{++}$ component depends only on $\xi^+$, and is called the
left-moving current, whereas $T_{--}$ depends only on $\xi^-$ and
is referred to as right-moving current. To ensure conformal
symmetry on the boundary, we need to impose boundary conditions on
the currents \rf{brane6}. In general, we find the boundary
condition for a given current  by using its associated charge.
Applied to the stress tensor, the result is
\begin{equation}\label{brane7}
T_{++}-T_{--}\;=\;0.
\end{equation}
Now, using the Eqs. \rf{brane2}, \rf{brane6} and \rf{brane7} we
find
\begin{equation}\label{brane8}
(-1)^{\Omega}{({\cal{R}}^{st})_{_{{\Upsilon}}}}^{^{\hspace{-0.5mm}\Omega}}\;
{ G}_{_{ \Omega\; \Xi}}\;{{\cal
R}^{^{\Xi}}}_{{\;\;\Lambda}}\;=\;G_{\Upsilon\Lambda}.
\end{equation}
Thus, in this way, we have derived the condition for conformal
invariance on the boundary in a sigma model on supermanifold. In
the following, we define a Dirichlet projector
${\cal{Q}}^{\Upsilon}_{\;\;\Lambda}$ on the worldsheet boundary,
which projects vectors onto the space normal to the brane. These
vectors (Dirichlet vectors) are eigenvectors of ${{\cal
R}^{\mu}}_{\nu} \left( {{\cal R}^{\alpha}}_{\beta} \right)$ with
eigenvalue $-1$. Hence ${\cal{Q}}^{\Upsilon}_{\;\;\Lambda}$ is
given by the following axioms
$$
{\cal{Q}}^2:={\cal{Q}}^{\Upsilon}_{\;\;\;\Xi}\;{\cal{Q}}^{\Xi}_{\;\;\Lambda}={\cal{Q}}^{\Upsilon}_{\;\;\Lambda},
$$
\begin{equation}\label{brane9}
{\cal{Q}}^{^{\Upsilon}}_{_{\;\;\Xi}}\;{{\cal
R}^{^{\Xi}}}_{{\;\Lambda}}\;=\;{{\cal
R}^{^{\Upsilon}}}_{_{\Xi}}\;{\cal{Q}}^{^{\Xi}}_{_{\;\;\Lambda}}=-{\cal{Q}}^{\Upsilon}_{\;\;\Lambda}.
\end{equation}
Similarly, we may define a Neumann projector ${\cal
N}^{\Upsilon}_{\;\;\Lambda}$ which projects vectors onto the
target space of the brane (vectors target to the brane are
eigenvectors of ${{\cal R}^{\mu}}_{\nu} \left( {{\cal
R}^{\alpha}}_{\beta} \right)$ with eigenvalue $1$) and is defined
as complementary to ${\cal{Q}}^{\Upsilon}_{\;\;\Lambda}$, i.e.,
\begin{equation}\label{brane10}
{\cal N}^{\Upsilon}_{\;~\Lambda}:={\delta}^{\Upsilon}_{\;~
\Lambda}-{\cal{Q}}^{\Upsilon}_{\;\;\Lambda}.
\end{equation}
In addition, by contracting \rf{brane10} with
${\cal{Q}}^{\Lambda}_{\;\;\Xi}$ and using \rf{brane9}, we then
obtain
\begin{equation}\label{brane11}
{\cal
N}^{^{\Upsilon}}_{_{\;\;\Lambda}}\;{\cal{Q}}^{^{\Lambda}}_{_{\;\;\Xi}}
=0.
\end{equation}
Also, the Neumann projector satisfy the following conditions
\begin{equation}\label{brane12}
(-1)^{\Omega}({\cal N}^{st})_{_{\Upsilon}}^{^{\;\;\Omega}}\;
({\cal E}^{st})_{_{{\Omega} \; \Xi}}\;{\cal
N}^{\Xi}_{\;~\Lambda}-(-1)^{\Xi} ({\cal
N}^{st})_{_{\Upsilon}}^{^{\;\;\Xi}}\; {\cal E}_{_{
\Xi\;\Omega}}\;{\cal{N}}^{^{\Omega}}_{_{\;\;\Delta}}\;{{\cal
R}^{^{\Delta}}}_{{\Lambda}}\;=\;0,~~~~~~~~~~~~~~~~~~~
\end{equation}
\begin{equation}\label{brane12.1}
~~~~~~~~~~~~~~~~(-1)^{\Omega}{({\cal{N}}^{st})_{_{{\Upsilon}}}}^{^{\hspace{-0.5mm}\Omega}}\;
{ G}_{_{\Omega \; \Xi}}\;{{\cal Q}^{^{\Xi}}}_{{\;\Lambda}}\;=\;0,
\end{equation}
note that for a spacefilling brane (when all directions are
Neumann or ${{\cal Q}^{^{\Upsilon}}}_{{\;\Lambda}}=0$) Eq.
\rf{brane12} implies that ${{\cal
R}^{^{\Upsilon}}}_{{\;\Xi}}=(-1)^{\Lambda}\;{({\cal
E}^{-1})^{\Upsilon\Lambda}}\;({{\cal E}^{st})_{\Lambda\Xi}}$.

Before proceeding  to discuss the dual conditions, let us write
down the boundary conditions \rf{brane2}, \rf{brane8},
\rf{brane9}, \rf{brane12} and \rf{brane12.1} in the Lie
superalgebra frame. These conditions read
\begin{equation}\label{brane13}
\hspace{5cm}~~~~~~~~~~~{R^{(l)}_-}^i\;=\;{{\cal
R}^{^{i}}}_{{\;j}}\;{R^{(l)}_+}^j,
\end{equation}
\begin{equation}\label{brane14}
~~~~~~~~~~~~~~~~~~(-1)^k
{({\cal{R}}^{st})_{_{{i}}}}^{^{\hspace{-0.5mm}k}}\;{ {
\Omega}}_{_{k l}}\;{{\cal R}^{^{l}}}_{{\;\;j}}\;=\;\Omega_{ij},
\end{equation}
\begin{equation}\label{brane15}
~~~~~~~~~~~~~~~~~~~~~{\cal{Q}}^{i}_{\;\;j}\;{{\cal
R}^{^{j}}}_{{\;k}}\;=\;{{\cal
R}^{^{i}}}_{{\;j}}\;{\cal{Q}}^{j}_{\;\;k}=-{\cal{Q}}^{i}_{\;\;k},
\end{equation}
\begin{equation}\label{brane16}
(-1)^k ({\cal N}^{st})_{i}^{\;~k}\;{({ F^+}^{st})}_{kl}\;{\cal
N}^{l}_{\;~j}-(-1)^l({\cal
N}^{st})_{i}^{\;~l}\;{F}_{lk}^+\;{\cal{N}}^{k}_{\;\;m}\;{{{\cal
R}}^{^{m}}}_{{j}}\;=\;0,~~~~~~~~~~~~~~~~~~~~~~~~
\end{equation}
\begin{equation}\label{brane16.1}
~~~~~~~~~~~~~~~~(-1)^j{({\cal{N}}^{st})_{_{{i}}}}^{^{\hspace{-0.5mm}j}}\;
{ \Omega}_{_{ j k}}\;{{\cal Q}^{^{k}}}_{{\;\;l}}\;=\;0,
\end{equation}
where
\begin{equation}\label{brane16.1.1}
~~{{ \cal R}^{^{i}}}_{{\;j}}=({R^{(l)}}^{st})^{i}_{\;\;\Upsilon}\;
{{\cal R}^{^{\Upsilon}}}_{{\;\Lambda}}\;
({R^{(l)}}^{-st})^{\Lambda}_{\;\;j},\;\;\;\;\;{\Omega}_{ij}=(-1)^{\Upsilon}
({R^{(l)}}^{-1})_{i}^{\;\;\Upsilon}\; { G}_{_{\Upsilon \Lambda}}
\; ({R^{(l)}}^{-st})^{\Lambda}_{\;\;j},
\end{equation}
\begin{equation}
{{\cal N}^{^{i}}}_{{\;j}}=({R^{(l)}}^{st})^{i}_{\;\;\Upsilon}\;
{{\cal N}^{^{\Upsilon}}}_{{\;\Lambda}}\;
({R^{(l)}}^{-st})^{\Lambda}_{\;\;j},\;\;\;\;\;\;{{\cal
Q}^{^{i}}}_{{\;j}}=({R^{(l)}}^{st})^{i}_{\;\;\Upsilon}\; {{\cal
Q}^{^{\Upsilon}}}_{{\;\Lambda}}\;
({R^{(l)}}^{-st})^{\Lambda}_{\;\;j}.~~~~~
\end{equation}
Note that the object ${{\cal R}^{^{i}}}_{{\;j}}$ is a gluing map
between currents at the worldsheet boundary. Since it maps
${R^{(l)}_+}^j$ to ${R^{(l)}_-}^i$, which are elements of the Lie
superalgebra, it is clearly a map from the Lie superalgebra into
itself. So, it may be assumed  to be a constant Lie superalgebra
automorphism, i.e.,  it preserves the Lie superalgebra structure.

\subsection{ {\it Supercanonical transformations}}

In this subsection, by using the super Poisson-Lie T-duality
transformation as a supercanonical transformation, we derive a
duality map for the gluing supermatrix which locally defines the
properties of the $D$-brane. In \cite{sfetsos1}, Sfetsos
formulated Poisson-Lie T-duality as an explicit transformation
between the canonical variables of the two dual sigma models.
Here, we generalize the classical canonical transformation on Lie
group \cite{{sfetsos1},{sfetsos2}} to the Lie supergroup. This
transformation on the Lie supergroup $G$ between the
supercanonical pairs of variables $({R^{(l)}_{\sigma}}^i
\;,\;P_i)$ and $(({{\tilde R}^{(l)}}_{\sigma})_j \;,\;{\tilde
P}^j)$ is given by
\begin{equation}\label{canonic1}
~~~~~~~~~~~~~~~~~~~~~~~{R^{(l)}_{\sigma}}^i\;=\;\left(
\delta^i_{\;j}- (-1)^k \Pi^{ik} {\tilde \Pi}_{kj}\right){\tilde
P}^j-(-1)^k \Pi^{ik}\;({{\tilde R}^{(l)}}_{\sigma})_{k},
\end{equation}
\begin{equation}\label{canonic2}
~~~~~~~~~~~~~~~~P_i\;=\; {\tilde \Pi}_{ij}{\tilde P}^j+({{\tilde
R}^{(l)}}_{\sigma})_{i},\hspace{3cm}~~~~~~~~~~~
\end{equation}
where\footnote{For calculating $P_\Upsilon$ in \rf{canonic4}, we
use the Lagrangian \rf{sigma12}. }
\begin{equation}\label{canonic3}
~{R^{(l)}_{\sigma}}^i\;=\; \frac{1}{2}\left( {R^{(l)}_{+}}^i
-{R^{(l)}_{-}}^i \right),\hspace{2cm}~~~
\end{equation}
$$
~P_i\;=\;
(-1)^{\Upsilon}\;({R^{(l)}}^{-1})_{i}^{\;\;\Upsilon}\;P_\Upsilon=(-1)^{\Upsilon}\;({R^{(l)}}^{-1})_{i}^{\;\;\Upsilon}\;
\frac{L\overleftarrow{\delta}}{\delta(\partial_{\tau}\Phi^{\Upsilon})}\hspace{5cm}
$$
\begin{equation}\label{canonic4}
~~~~=\frac{1}{2}\left(({F^{+^{st}}})_{ij}\;{R^{(l)}_{+}}^j+F^+_{ij}{R^{(l)}_{-}}^j
\right),
\end{equation}
and similarly for the corresponding tilded  symbols. To find the
dual boundary conditions, one must find a transformation from
${R^{(l)}_{\pm}}$ to ${{\tilde R}^{(l)}_{\pm}}$. For this purpose,
we use   Eqs. \rf{sigma15}, \rf{sigma21}, \rf{canonic3} and
\rf{canonic4} to rewrite the supercanonical transformations
\rf{canonic1} and \rf{canonic2} as follows:
\begin{equation}\label{canonic5}
~~~({{{\tilde R}^{(l)}_{+}}})_i\;=\;(-1)^l\;({{\tilde
F}^{+^{-st}}})_{ij}({{ E}^{+^{-st}}})^{jl}(e)({{
F}^{+^{st}}})_{lk}\;{R^{(l)}_{+}}^k,
\end{equation}
\begin{equation}\label{canonic6}
({{{\tilde R}^{(l)}_{-}}})_i\;=-\;(-1)^l\;({{\tilde
F}^{+^{-1}}})_{ij}({{ E}^{+^{-1}}})^{jl}(e){{
F^+}}_{lk}\;{R^{(l)}_{-}}^k.
\end{equation}
Now, by using the above relations, the boundary condition
\rf{brane13} takes the following form\footnote{Here in the Lie
superalgebra frame, ${{{\tilde {\cal
R}}}^{^{\Upsilon}}}{_{_{\Lambda}}}$ reads

$$
{{{\tilde {\cal R}}}_i}^{\;\;j}\;=\;({{\tilde R}^{(l)^{st}}})_{i
\Upsilon} \;{{{{\tilde {\cal
R}}}^{^{\Upsilon}}}{_{_{\Lambda}}}}\;({{\tilde
R}^{(l)^{-st}}})^{\Lambda j}.
$$}
\begin{equation}\label{canonic7}
({{{\tilde R}^{(l)}_{-}}})_i\;=\;(-1)^j\;{{\tilde {\cal
R}}_i}^{\;\;j}\;({{{\tilde R}^{(l)}_{+}}})_j,
\end{equation}
in which
\begin{equation}\label{canonic8}
{{\tilde {\cal R}}_i}^{\;\;j}\;=\;-(-1)^{l+p}\;({{\tilde
F}^{+^{-1}}})_{ik}({{ E}^{+^{-1}}})^{kl}(e){ F^+}_{lm}\;{{\cal
R}^{m}_{\;\;n}}\;({{ F}^{+^{-st}}})^{np}({{
E}^{+^{st}}})_{pq}(e)({{\tilde F}^{+^{st}}})^{qj}.
\end{equation}
The above relation is the transformation of the gluing
supermatrix. By using the \rf{canonic8} and the rules of
supertranspose \cite{D}  we have $sdet({_i{\tilde {\cal
R}}}^j)=sdet(-{\cal R}^i_{\;\;j})$. This is a result that will be
useful in the next subsection. Furthermore, again by use of the
\rf{canonic8} one can determine the form of the dual Neumann and
Dirichlet projectors ${\tilde {\cal N}}$ and ${\tilde {\cal Q}}$
via the definition $(-1)^{j}{{\tilde {\cal R}}_i}^{\;\;j}{{\tilde
{\cal Q}}_j}^{\;\;k}=(-1)^{j}{{\tilde {\cal Q}}_i}^{\;\;j}{{\tilde
{\cal R}}_j}^{\;\;k}=-{{\tilde {\cal Q}}_i}^{\;\;k}$ and so on.

Similarly, to obtain the transformation of the metric on Lie
superalgebra we use the relation \rf{brane14}. Thus, the dual of
the \rf{brane14} is found  to be
\begin{equation}\label{canonic9}
(-1)^l\;{({\tilde {\cal R}}^{st})^{i}}_{\;k}\;{{\tilde {
\Omega}}}^{kl}\;{{{\tilde {\cal R}}}_l}^{\;\;j}\;=\; {{\tilde {
\Omega}}}^{ij},
\end{equation}
where
\begin{equation}\label{canonic10}
{{\tilde { \Omega}}}^{ij}\;=\;(-1)^{l+n+p}\;({{\tilde
F^+}}){^{il}}{{ E^+}_{\hspace{-2mm}lm}}(e){( F^{+^{-1}})}^{mn}\;{
\Omega}_{nk}\;({{ F}^{+^{-st}}})^{kp}({{
E}^{+^{st}}})_{pq}(e)({{\tilde F}^{+^{st}}})^{qj}.
\end{equation}

\subsection{ {\it Example}}

We now investigate the consequences of the duality transformation
of the gluing supermatrix for the Drinfel'd superdouble
$(gl(1|1)\;,\; {{\cal B} \oplus {\cal A } \oplus {\cal
A}_{1,1}}_|. i)$. The (anti)commutation relations of this
superdouble and the super Poisson-Lie T-dual sigma models have
been explicitly worked out  in section 4. The constant background
at the identity as the relation \rf{Lie4} and the super Poisson
brackets have been given by the relations \rf{Lie3} and
\rf{Lie28}. Thus, using the relations \rf{sigma15}, \rf{sigma21}
and \rf{sigma22}, the background fields ${F}_{ij}^{+}(g)$ and
${{\tilde F}^{+ij}}(\tilde g)$ read
\begin{equation}\label{ex1}
{F}_{ij}^{+}(g) \;=\;\pmatrix{0 & -1 & 0 & e^{y} {\psi} \cr -1 & 0
& 0 & 0 \cr 0 & 0 & 0 & -1 \cr -e^{y} {\psi}  & 0 & 1 & 0 },
\end{equation}
\begin{equation}\label{ex2}
{{\tilde F}^{+ij}}(\tilde g) \;=\;\pmatrix{0 & -1 & 0 & 0 \cr -1 &
\frac{-2\tilde \psi \tilde \chi}{e^{\tilde x}-2} & \tilde \chi
e^{-\tilde x}& \frac{-\tilde \psi e^{\tilde x}}{e^{\tilde x}-2}
\cr 0 & \frac{\tilde \chi }{e^{\tilde x}-2} & 0 &
\frac{1}{e^{\tilde x}-2} \cr 0  & -\tilde \psi & e^{-\tilde x} & 0
}.
\end{equation}
In this example (a supergroup  with $(2|2)$-dimension), we have
the following six different types of $D$-branes where for each of
these
cases we find the dual gluing supermatrix.\\

{\bf Case~1:} Case 1 refers to $D_{(-1)|0}$-brane. The
corresponding gluing supermatrix is given by
\begin{equation}\label{ex3}
{\cal R}^i_{\;j} \;=\;\pmatrix{-1 & 0 & 0 & 0 \cr 0 & -1 & 0 & 0
\cr 0 & 0 & -1 & 0 \cr 0  & 0 & 0 & -1 }.
\end{equation}
This choice means that all directions are Dirichlet, i.e.,
${\cal{Q}}^{i}_{\;\;j}={\delta}^{i}_{\;\;j}$ and
${\cal{N}}^{i}_{\;\;j}=0$. Then, using the relation \rf{Lie4} and
by substituting \rf{ex1} and \rf{ex2} into  \rf{canonic8}, the
dual gluing supermatrix reads
\begin{equation}\label{ex4}
_i{{\tilde {\cal R}}}^{\;j}\;=\; \pmatrix{1 & {\tilde A}_1 &
-\frac{2\tilde \chi+2\psi e^y}{e^{\tilde x}-2} & 2{\tilde \psi}
\cr 0 & 1 & 0& 0 \cr 0 & \frac{-2\tilde \psi e^{\tilde x}
}{e^{\tilde x}-2} & \frac{-e^{\tilde x}}{e^{\tilde x}-2} & 0 \cr 0
& {\tilde A}_2 & 0 & 2e^{-\tilde x}-1 },
\end{equation}
where
$$
{\tilde A}_1=\frac{2}{e^{\tilde x}-2}\left[ \psi {\tilde \psi}
e^{\tilde x+y}(e^{\tilde x}-3)+2{\tilde \psi}{\tilde \chi}
\right],\;\;\;\;\;\;\;{\tilde A}_2=2\psi e^{y} ({e^{\tilde x}-2})
-2{\tilde \chi}e^{-\tilde x}.
$$
The superdeterminant is $sdet({_i{\tilde {\cal R}}}^j)=sdet(-{\cal
R}^i_{\;\;j})=1$, so the dual brane may include the following
directions, for some special  backgrounds

$(i)~$ $D_{(-1)|0}$-brane, with the Neumann projector $_i{{\tilde
{\cal N}}}^{\;j}=0$,

$(ii)~$ $D_{(-1)|2}$-brane, with   $_i{{\tilde {\cal
N}}}^{\;j}$=diag(0, 0, 1, 1),

$(iii)~$ $D_{1|0}$-brane, with   $_i{{\tilde {\cal
N}}}^{\;j}$=diag(1, 1, 0, 0),

$(iv)~$ $D_{1|2}$-brane (spacefilling), with   $_i{{\tilde {\cal
N}}}^{\;j}=_i{\delta}^{\;j}$.\\
For the subcase $(i)$, the only solution is $_i{{\tilde {\cal
R}}}^{\;j}=-_i{\delta}^{\;j}$, i.e., all directions are Dirichlet,
such that Eq. \rf{canonic8} reduces to $(-1)^n {\tilde
F}^{+^{in}} ({{ \tilde F}^{+^{-st}}})_{nj}=-(-1)^{k+m}({{
E}^{+^{-1}}})^{ik}{ F}_{kl}^+\;({{ F}^{+^{-st}}})^{lm}\\({{
E}^{+^{st}}})_{mj}$. For the subcase $(ii)$, the dual brane has
two Dirichlet directions for the bosonic part and two Neumann
directions for the fermionic one. In contrast to the subcase
$(ii)$, in the subcase $(iii)$, the dual brane has zero Dirichlet
directions for the bosonic part and two Dirichlet directions for
the fermionic one. In the latter subcase, the only solution is
$_i{{\tilde {\cal R}}}^{\;j}=_i{\delta}^{\;j}$, i.e., the dual
brane has zero Dirichlet directions and  the relation
\rf{canonic8} reduces to $\delta^i_{~j}=(-1)^{k+m}\;({{
E}^{+^{-1}}})^{ik}{ F}_{kl}^+\;({{ F}^{+^{-st}}})^{lm}({{
E}^{+^{st}}})_{mj}$,  hence we find $\Pi^{ij}(g)=0$, i.e., for
this subcase we must have super non-Abelian $T$-duality.\\

{\bf Case~2:} In this  case, the corresponding gluing supermatrix
is given by
\begin{equation}\label{ex5}
{\cal R}^i_{\;j} \;=\;\pmatrix{-1 & 0 & 0 & 0 \cr 0 & -1 & 0 & 0
\cr 0 & 0 & 1 & 0 \cr 0  & 0 & 0 & 1 }.
\end{equation}
This is a $D_{(-1)|2}$-brane, with two Dirichlet directions for
the bosonic part and two Neumann directions for the fermionic one.
The dual gluing supermatrix again follows from \rf{canonic8}:
\begin{equation}\label{ex6}
_i{{\tilde {\cal R}}}^{\;j}\;=\; \pmatrix{1 & 0 & 0 & 0 \cr 0 & 1
& 0& 0 \cr 0 & \frac{2\tilde \psi e^{\tilde x}(e^{\tilde x}-1)
}{e^{\tilde x}-2}& \frac{e^{\tilde x}}{e^{\tilde x}-2} & 0 \cr 0 &
2\tilde \chi (e^{-\tilde x}-1)  & 0 & 1-2e^{-\tilde x} }.
\end{equation}
The superdeterminant is $sdet({_i{\tilde {\cal R}}}^j)=1$. In this
case,  the dual branes may include the same  directions of the
dual branes in Case 1.\\

{\bf Case~3:} In this case, we have a $D_{0|0}$-brane, with the
following gluing supermatrix
\begin{equation}\label{ex6.1}
{\cal R}^i_{\;j} \;=\;\pmatrix{1 & 0 & 0 & 0 \cr 0 & -1 & 0 & 0
\cr 0 & 0 & -1 & 0 \cr 0  & 0 & 0 & -1 },
\end{equation}
with one Dirichlet direction  and one Neumann direction for the
bosonic part and two Dirichlet directions for the fermionic one.
Then Eq. \rf{canonic8} yields the dual gluing supermatrix
\begin{equation}\label{ex7}
_i{{\tilde {\cal R}}}^{\;j}\;=\; \pmatrix{1 & \frac{2\tilde \psi}{e^{\tilde x}-2}(\psi
e^{\tilde x+y} +2 \tilde \chi ) & \frac{-2(\tilde \chi+ \psi
e^{y}) }{e^{\tilde x}-2} & 2 \tilde \psi \cr 0 & -1 & 0& 0 \cr 0 &
\frac{-2\tilde \psi e^{\tilde x}(e^{\tilde x}-1) }{e^{\tilde
x}-2}& \frac{-e^{\tilde x}}{e^{\tilde x}-2} & 0 \cr 0 & 2\tilde
\chi (1-e^{-\tilde x})  & 0 & 2e^{-\tilde x}-1 }.
\end{equation}
The superdeterminant is $sdet({_i{\tilde {\cal R}}}^j)=-1$, so it
is either a $D_{0|0}$-brane or a $D_{0|2}$-brane.\\

{\bf Case~4:} This case introduces  the following gluing
supermatrix
\begin{equation}\label{ex6.1}
{\cal R}^i_{\;j} \;=\;\pmatrix{1 & 0 & 0 & 0 \cr 0 & -1 & 0 & 0
\cr 0 & 0 & 1 & 0 \cr 0  & 0 & 0 & 1 }.
\end{equation}
This is a $D_{0|2}$-brane, with one Dirichlet direction and one
Neumann direction for the bosonic part and zero Dirichlet
directions for the fermionic one. The dual gluing supermatrix
reads
\begin{equation}\label{ex8}
_i{{\tilde {\cal R}}}^{\;j}\;=\; \pmatrix{1 & 2{\tilde \psi }\psi
e^{\tilde x+y} & 0 & 0 \cr 0 & -1 & 0& 0 \cr 0 &
\frac{2\tilde \psi e^{\tilde x} }{e^{\tilde x}-2}& \frac{e^{\tilde
x}}{e^{\tilde x}-2} & 0 \cr 0 & -{\tilde A}_2 & 0 & 1-2e^{-\tilde
x} }.
\end{equation}
Its superdeterminant is $-1$, so, the dual branes are
$D_{0|0}$-brane or  $D_{0|2}$-brane.\\

{\bf Case~5:} The corresponding gluing supermatrix for this case
has the following form
\begin{equation}\label{ex9}
{\cal R}^i_{\;j} \;=\;\pmatrix{1 & 0 & 0 & 0 \cr 0 & 1 & 0 & 0 \cr
0 & 0 & -1 & 0 \cr 0  & 0 & 0 & -1 }.
\end{equation}
This is a $D_{1|0}$-brane, with zero Dirichlet direction  for the
bosonic part and two Dirichlet directions for the fermionic one.
The dual gluing supermatrix becomes
\begin{equation}\label{ex10}
_i{{\tilde {\cal R}}}^{\;j}\;=\; \pmatrix{-1 & 0 & 0 & 0 \cr 0 &
-1 & 0& 0 \cr 0 & \frac{-2\tilde \psi e^{\tilde x}(e^{\tilde x}-1)
}{e^{\tilde x}-2}& \frac{-e^{\tilde x}}{e^{\tilde x}-2} & 0 \cr 0
& 2\tilde \chi (1-e^{-\tilde x}) & 0 & 2e^{-\tilde x} -1}.
\end{equation}
It has superdeterminant $sdet({_i{\tilde {\cal R}}}^j)=1$, so the
dual branes may include the same  directions of the dual branes in
Case 1.\\

{\bf Case~6:} This case is devoted to a spacefilling $D$-brane,
i.e., $D_{1|2}$-brane. The corresponding gluing supermatrix,
according to Eq. \rf{brane16} is given by
\begin{equation}\label{ex11}
{{\cal R}^{i}}_{{\;j}}\;=\;(-1)^{k}\;{({ F}^{+^{-1}})^{ik}}\;({{
F}^{+^{st}})_{kj}}\;=\; \pmatrix{1 & 0 & 0 & 0 \cr 0 & 1 & 0&
2\psi e^{y} \cr 2\psi e^{y} & 0 & 1 & 0 \cr 0 & 0 & 0 & 1}.
\end{equation}
Then, the dual gluing supermatrix is found to be of the form
\begin{equation}\label{ex12}
_i{{\tilde {\cal R}}}^{\;j}\;=\; \pmatrix{-1 & \frac{-4\tilde \psi \tilde \chi}{e^{\tilde x}-2} & \frac{2\tilde \chi
}{e^{\tilde x}-2} & -2{\tilde \psi} \cr 0 &
-1 & 0& 0 \cr 0 & \frac{2\tilde \psi e^{\tilde x}
}{e^{\tilde x}-2}& \frac{e^{\tilde x}}{e^{\tilde x}-2} & 0 \cr 0
& 2e^{-\tilde x} \tilde \chi  & 0 & 1-2e^{-\tilde x} }.
\end{equation}
Its superdeterminant is 1, so for the dual branes we have the same
directions of the dual branes in Case 1.

In this example, we showed how  $D$-branes in the model are
exchanged. At the end, we shall discuss the boundary conditions
for  Case 2. First, by insertion of relations \rf{Lie2} into
\rf{sigma13}, we compute the  ${{_{_\Upsilon}} R}^{(l)^i}$. Then,
by substituting  \rf{ex5} into the first equation of
\rf{brane16.1.1}, we obtain
\begin{equation}\label{ex13}
{{\cal R}^{\Upsilon}}_{{\;\Lambda}}\;=\; \pmatrix{-1 & 0 & 0 & 0 \cr 0 & -1 & 2 e^{y} \chi&
0 \cr 0 & 0 & 1 & 0 \cr 2\chi & 0 & 0 & 1}.
\end{equation}
In terms of the ${{\cal R}^{\Upsilon}}_{{\;\Lambda}}$, the gluing
condition \rf{brane2} implies the Dirichlet boundary condition
$\partial_{\tau}y=0$ for the field $y$. The boundary conditions
for the remaining fields are of the form
\begin{equation}\label{ex14}
\partial_{\tau}x=e^y \chi\; \partial_{\tau}{\psi},~~~~~~~~~~~~\partial_{\sigma}{\chi}=- \chi\;
\partial_{\sigma}{y},~~~~~~~~~~~~~~~\partial_{\sigma}{\psi}=0.
\end{equation}
By imposing the latter condition on  the first two equations, we
then obtain $\partial_{\tau}\partial_{\sigma}x=0$. i.e.,
$\partial_{\tau}x=f(\tau)$ or $\partial_{\sigma}x=g(\sigma)$.
Similarly, using the \rf{Lie7-1} and \rf{ex6}, ${{{{\tilde {\cal
R}}}^{^{\Upsilon}}}{_{_{\Lambda}}}}$ reads
\begin{equation}\label{ex15}
{{{{\tilde {\cal R}}}^{^{\Upsilon}}}{_{_{\Lambda}}}}\;=\;
\pmatrix{1 & 0 & 0 & 0 \cr 0 & 1 & 0& 0 \cr 0 & -\frac{2\tilde
\psi (e^{\tilde x}-1) }{e^{\tilde x}-2}& \frac{e^{\tilde
x}}{e^{\tilde x}-2} & 0 \cr 0 & 2\tilde \chi (1-e^{-\tilde x})  &
0 & 1-2e^{-\tilde x} }.
\end{equation}
Now, by employing \rf{brane2}, the dual boundary conditions are
found to be
$$
~~~~~~~~~~~~~~~~~~~~~~~~~~\partial_{\sigma}{\tilde y}=0,
$$
$$
~~~~~~~~~~~~~~~~~~~~~~~~~~\partial_{\sigma}{\tilde x}=0,
$$
$$
-{\tilde \psi} (e^{\tilde x}-1)\partial_{\tau}{\tilde
x}+(e^{\tilde x}-1)\partial_{\sigma}{\tilde
\psi}+\partial_{\tau}{\tilde \psi}=0,~~~~~~~~~~~~~~~~
$$
\begin{equation}\label{ex16}
{\tilde \chi} (e^{\tilde x}-1)\partial_{\tau}{\tilde x}+(e^{\tilde
x}-1)\partial_{\sigma}{\tilde \chi}-\partial_{\tau}{\tilde
\chi}=0.~~~~~~~~~~~~~
\end{equation}
Note that the latter two equations imply the following condition
\begin{equation}\label{ex17}
(e^{\tilde x}-1)\partial_{\sigma}({\tilde \chi} {\tilde
\psi})+{\tilde \chi}\partial_{\tau}{\tilde \psi}+ {\tilde \psi}
\partial_{\tau}{\tilde \chi}=0.
\end{equation}


\newsection{ {\bf Conclusion}}
We have proved that the WZNW model on the Lie supergroup $GL(1|1)$ has
super Poisson-Lie symmetry with the dual Lie supergroup ${B \oplus
A \oplus A_{1,1}}_|.i$. Then, we discussed about $D$-branes and
worldsheet boundary conditions on supermanifolds, in general, and
obtained the algebraic relations on the gluing supermatrix for the
Lie supergroup case. Also, using the supercanonical
transformation description  of the super Poisson-Lie T-duality
transformation, we obtained formulae for the description of the dual
gluing supermatrix, then, we found the gluing supermatrix for the
WZNW model on  $GL(1|1)$ and its dual model. In this way, there are some new perspectives
to find super Poisson-Lie symmetry on the superstring models on Ads backgrounds. Investigation of superconformal boundary conditions \cite{L} on supermanifolds may become another open problem. Some of these
open problems are under current investigation.

\vspace{5mm}

 {\it Acknowledgment:~}This research was supported by a research fund No. 401.231 from Azarbaijan Shahid Madani University.
 We would like to thank F. Darabi for carefully reading the
manuscript and useful comments.


\end{document}